\documentclass[12pt]{iopart}
\usepackage{iopams} 
\usepackage{color}
\usepackage{graphicx} 
\usepackage{subfigure} 
\usepackage{tikz} 
\usepackage{epstopdf} 
\usepackage{cite}

\begin{document}

\title[Thermal ratchet effect in confining geometries]
{Thermal ratchet effect in confining geometries}

\author{Viktor Holubec $^{1}$, Artem Ryabov $^{1,\dag}$, Mohammad Hassan Yaghoubi $^{2,3}$, Martin Varga $^{1}$, Ayub Khodaee $^{2,4}$, M.\ Ebrahim Foulaadvand $^{2}$ and Petr Chvosta $^{1}$}

\address{$^{1}$ Charles University, Faculty of Mathematics and Physics, Department of Macromolecular Physics, V~Hole{\v s}ovi{\v c}k{\' a}ch~2, 180~00~Praha, Czech Republic}
\address{$^{2}$ Department of Physics, University of Zanjan, P.O.~Box~45196-311, Zanjan, Iran}
\address{$^{3}$ Complexity Science Group, Department of Physics \& Astronomy, University of Calgary, Calgary, Alberta~T2N~1N4, Canada}
\address{$^{4}$ Bradley Plasma Lab, Department of Physics \& Engineering Physics, University of Saskatchewan, Saskatoon, Saskatchewan SK~S7N~5E2, Canada }

\ead{$^{\dag}$rjabov.a@gmail.com}

%\ead{yaghoobimh@yahoo.com} 
%\ead{Viktor.Holubec@mff.cuni.cz}
%\ead{Martin.Varga@mff.cuni.cz} 
%\ead{foolad@iasbs.ac.ir}
%\ead{Petr.Chvosta@mff.cuni.cz} 

\begin{abstract}
Stochastic model of the Feynman-Smoluchowski ratchet is proposed and solved using generalization of the Fick-Jacobs theory. The theory fully captures nonlinear response of the ratchet to the difference of heat bath temperatures. The ratchet performance is discussed using the mean velocity, the average heat flow between the two heat reservoirs and the figure of merit, which quantifies energetic cost for attaining a certain mean velocity. Limits of the theory are tested comparing its predictions to numerics. We also demonstrate connection between the ratchet effect emerging in the model and rotations of the probability current and explain direction of the mean velocity using simple discrete analogue of the model.
\end{abstract}

\maketitle
 
%%%%%%%%%%%%%%%%%%%%%%%%%%%%%%%%%%%%%%%%%%%%%%%%%%%%%%%%%%%%%%%%%%%%%%%%%%%%%%%%%%%%%%%%%%%%%%%%%%%%%%%%%%%%%%%%%%%%%%%%%%%%%%%%%%%
%%%%%%%%%%%%%%%%%%%%%%%%%%%%%%%%%%%%%%%%%%%%%%%%%%%%%%%%%%%%%%%%%%%%%%%%%%%%%%%%%%%%%%%%%%%%%%%%%%%%%%%%%%%%%%%%%%%%%%%%%%%%%%%%%%%
\section{Introduction}

Diffusion in narrow channels of varying cross-sections, e.g.\ through micropores of zeolites or channels in cell membranes, is essentially a three-dimensional (3D) problem with rather complex boundary conditions. An elegant method how to reduce dimensionality of the underlying diffusion equation was proposed by Jacobs \cite{bookJacobs}. The main idea is to separate longitudinal and transversal dynamics in the narrow channel. After the separation, the narrow segments of the channel act effectively as entropic free-energy barriers hindering the 1D longitudinal diffusional dynamics.

The elegant Fick-Jacobs theory regained significant attention after the influential works \cite{Zwanzig1992, RegueraRubi} were published. In these, and in the subsequent works, the originally phenomenological approach was generalized and put on solid mathematical grounds
\cite{KalinayPercus2005, KalinayPercusPRE2005, KalinayPercusPRE2006, BuradaPTRA2009, BuradaSchmid2009, Drazer2009, Leo2010JCP, LeoBerezh2010JCP, KalinaySoft2011, KalinayPRE2011, MartensPRE2011, MartensChaos2011, PinedaLeoJCP2011, Dagdug2012, DagdugPinedaJCP2012, Martens2012Communication, KalinayJCP2013, LeoJCP2013, Martens2013, MartensPRL2013, Metzler2014, KalinaySinusDrivingPRE2014, Leo2014PhysicaA, Leo2014PRE,  Drazer2015, DasRayPRE2015, Leo2015JCP, Wang2016, Leo2016CW, Malgaretti2016, Malgaretti2016Polymer, KalinayPRE2016, Kalinay2017}. In particular, various many-dimensional Brownian ratchets have been studied with the aid of the Fick-Jacobs approximation including flashing and rocking ratchets  \cite{ConfinedRubi2012, Makhnovskii2012, ConfinedRubi2013}, ratchets driven by a temperature gradient  \cite{ConfinedRubi2013}, and hydrodynamic ratchets \cite{Martens2013, Slanina2016PRE}. Possible application to separation of particles according to their size can be found in Refs.~\cite{SlaterElectrolytes97, RegueraPRL2012, PhysRevLett.88.168301, C2LC21089D}. 

In the present paper we exploit a generalization of the Fick-Jacobs theory proposed in our recent work \cite{RatchetJSTAT2016} to describe dynamics, energetics and performance of a stochastic Feynman-Smoluchowski ratchet \cite{Smoluchowski1912,Feynman}. 
The ratchet, used by Feynman as a pedagogical gedankenexperiment, provides insight into possible working principles of molecular machines \cite{Reimann, Erbas-Cakmak2015} and serves as one of the basic models of non-equilibrium stochastic energetics \cite{SekimotoBook, Seifert2012}. As such its several analogues have been intensively studied in recent years \cite{Reimann, SekimotoBook, Seifert2012, OonoPaniconi1998, HatanoSasa2001, Komatsu2010, Sekimoto1997, Magnasco1998, Hondou1998, JarzynskiWojcik2004, KomatsuPRE2006, Gomez-Marin2006, RoeckMaes2007, KomatsuPRL2008, Ueda2012, JarzynskiMazonka1999, Tumlin2016}, yet analytically solvable qualitative models are rather rare \cite{JarzynskiMazonka1999,  Visco2006, Fogedby2011, Dotsenko2013, Grosberg2015}. 

Advantage of the present approach over previous works is that it provides both analytical quantitative description (not restricted to linear-response regime) and a clear qualitative insight into fundamental working principles of the 2D thermal ratchet. The proposed model shares all essential features of the famous ratchet from Feynman's lectures (cf.\ Fig.~$2.1.$~in Ref.~\cite{Reimann} and Fig.~\ref{fig:model} below): randomly rotating asymmetric wheels and a pawl which should rectify the rotatory Brownian motion of the wheels; the both parts are coupled to reservoirs at different temperatures. Besides deriving analytical formulas we also emphasize connection between the ratchet effect emerging in the model and rotations of the probability current.

The article is organized as follows. The model is defined in Sec.~\ref{sec:model}, Sec.~\ref{sec:FP} comprises solution of the Fokker-Planck equation. In Sec.~\ref{sec:vQ} we discuss the mean velocity of rotating wheels, the mean heat flow between the reservoirs and the figure of merit of the ratchet. In Sec.~\ref{sec:circulation} we present an explanation of the ratchet effect based on the circulation of probability current. Direction of the circulation is justified using rough discrete analogue of the model.

%%%%%%%%%%%%%%%%%%%%%%%%%%%%%%%%%%%%%%%%%%%%%%%%%%%%%%%%%%%%%%%%%%%%%%%%%%%%%%%%%%%%%%%%%%%%%%%%%%%
%%%%%%%%%%%%%%%%%%%%%%%%%%%%%%%%%%%%%%%%%%%%%%%%%%%%%%%%%%%%%%%%%%%%%%%%%%%%%%%%%%%%%%%%%%%%%%%%%%%
\begin{figure}[t!]
\centering
\qquad
\subfigure{
\begin{tikzpicture}[scale=0.46,
      trans/.style={thick,shorten >=2pt,shorten <=2pt,>=stealth},
]
\draw [fill=white,thick, dashed, draw=red] (1.5,-7.0) rectangle (5.5,-3.5);
\draw [fill=white,thick, dashed, draw=blue] (-3,-2.6) rectangle (10.0,2.6);
\draw [fill=white,thick] (4.5,0.3) rectangle (2.5,0.1);
\draw [fill=white,thick] (3.55,0.1) rectangle (3.45,-6.0);
\draw [black,thick,<->,domain=30:-60] plot ({1*cos(\x)}, {1*sin(\x)});
\draw (0,0) [thick] circle (0.2cm);
\draw[trans,<->] (2.5,-4.5) -- (4.5,-4.5) node[midway,midway] {};
\node[draw=none] at (4.5,-5.5) {$T_y$};
\node[draw=none] at (3.4,1.9) {$T_x$};
\node[draw=none] at (1.4,0.2) {$x$};
\draw[thick]
\foreach \i in {1,2,...,10} {%
   [rotate=(\i-1)*36]  (0:2) arc (2.0:12:3) -- (20:2.3)  arc
(17.8:30:2.9) --  (36:2)
 };
\begin{scope}[xshift=7.0cm,rotate=180/8]
\draw [black,thick,<->,domain=7.5:-82.5] plot ({1*cos(\x)}, {1*sin(\x)});
\draw (0,0) [thick] circle (0.2cm);
\node[draw=none] at (1.4,-0.5) {$x$};
\draw[thick]
\foreach \i in {1,2,...,10} {%
   [rotate=(\i-1)*36]  (0:2) arc (2.0:12:3) -- (20:2.3)  arc
(17.8:30:2.9) --  (36:2)
 };\end{scope}
\node[draw=none] at (4.0,-4.0) {$y$};
\draw[fill=white,thick] (1.1,1.05) rectangle (8.1,1.15);
\draw [fill=white,thick] (1.1,1.1) circle (0.1cm);
\draw [fill=white,thick] (8.1,1.1) circle (0.1cm);
\end{tikzpicture}
}\qquad 
\subfigure{
\includegraphics[width=0.4\linewidth]{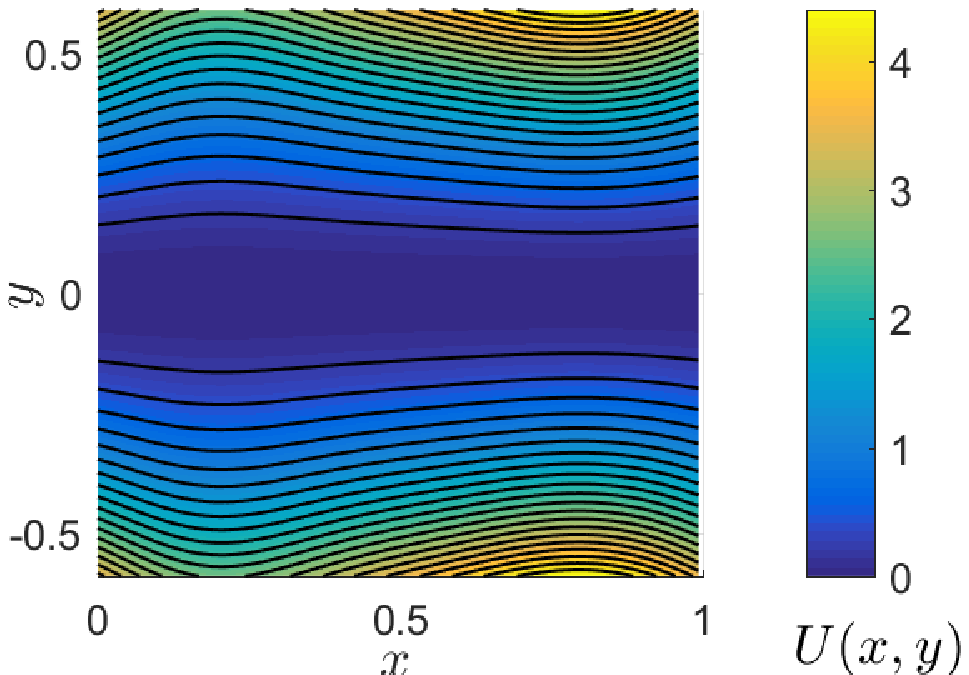}
}
\qquad 
\subfigure{
\includegraphics[width=0.4\linewidth]{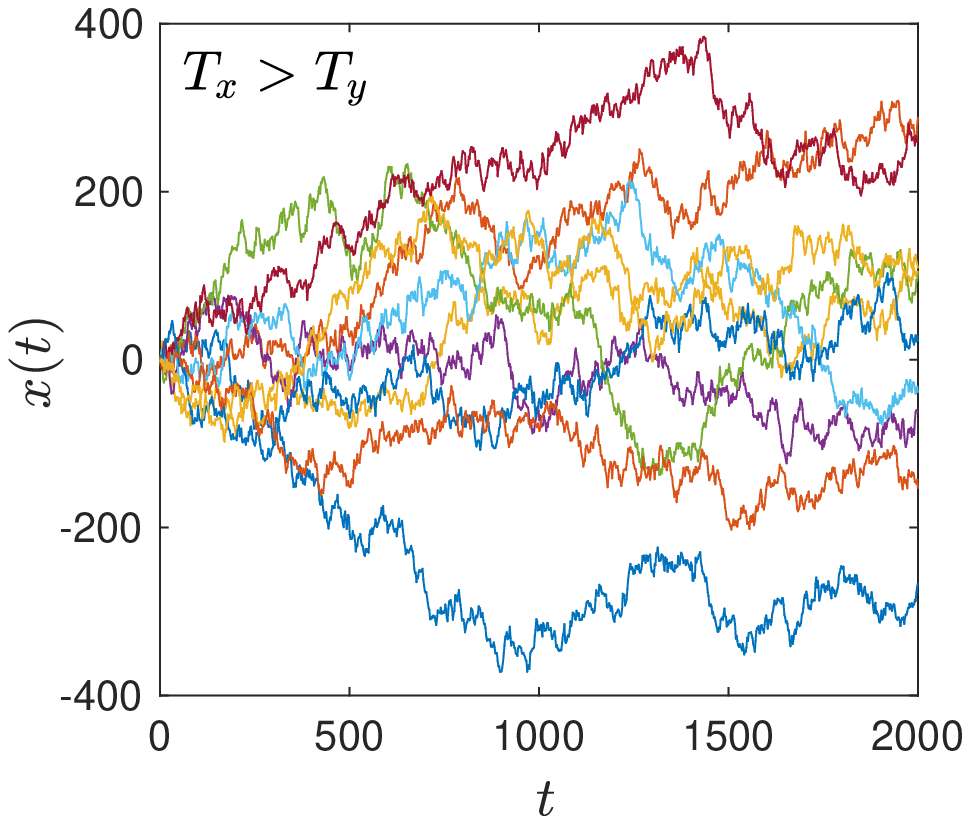}
}
\qquad 
\subfigure{
\includegraphics[width=0.4\linewidth]{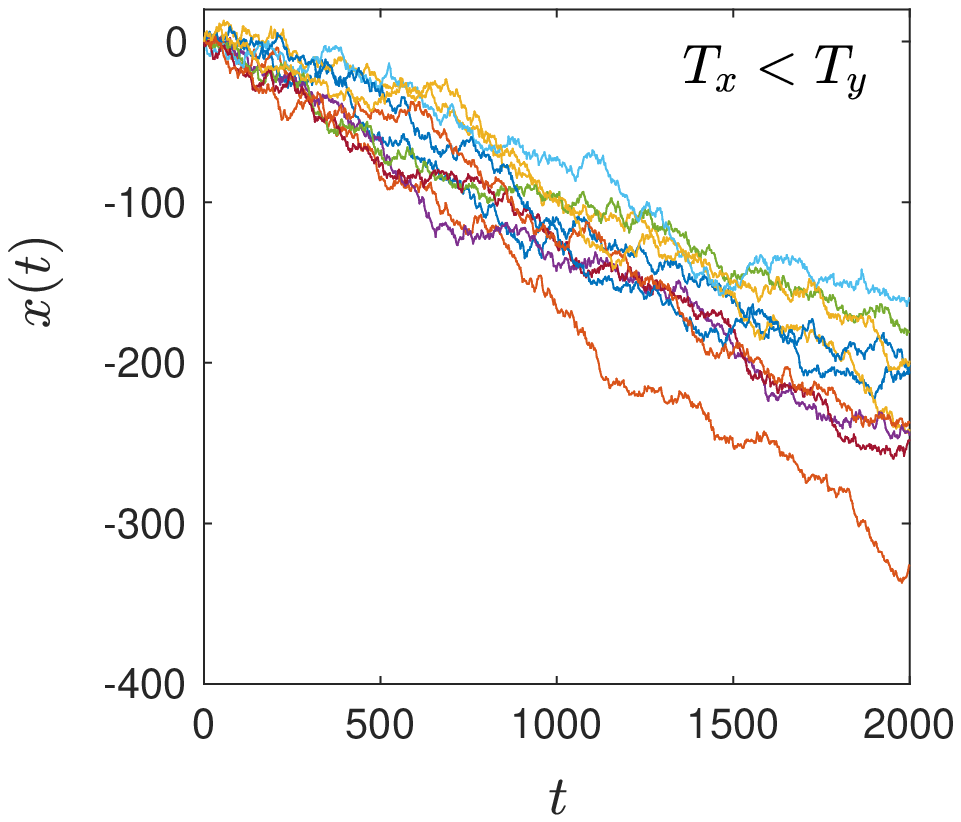}
}
\caption{Upper left: Schematic of the mechanical ratchet. Upper right: Equipotentials of the parabolic potential (\ref{potentialGEN}) representing interaction between wheels and the pawl. The spring stiffness $k(x)$ captures asymmetry of ratchet teeth, it is given in Eq.~(\ref{SpringConst}). Lower panels: few trajectories obtained by the numerical integration of the Langevin equations (\ref{Langevin}) using the Euler-Maruyama method \cite{bookKloeden}. We have used $T_x = 10$, $T_y = 1$ in the left panel and $T_x = 1$, $T_y = 10$ in the right panel ($\varepsilon^{2}=0.01$).}
\label{fig:model} 
\end{figure}
%%%%%%%%%%%%%%%%%%%%%%%%%%%%%%%%%%%%%%%%%%%%%%%%%%%%%%%%%%%%%%%%%%%%%%%%%%%%%%%%%%%%%%%%%%%%%%%%%%%
%%%%%%%%%%%%%%%%%%%%%%%%%%%%%%%%%%%%%%%%%%%%%%%%%%%%%%%%%%%%%%%%%%%%%%%%%%%%%%%%%%%%%%%%%%%%%%%%%%% 

%%%%%%%%%%%%%%%%%%%%%%%%%%%%%%%%%%%%%%%%%%%%%%%%%%%%%%%%%%%%%%%%%%%%%%%%%%%%%%%%%%%%%%%%%%%%%%%%%%%%%%%%%%%%%%%%%%%%%%%%%%%%%%%%%%%
%%%%%%%%%%%%%%%%%%%%%%%%%%%%%%%%%%%%%%%%%%%%%%%%%%%%%%%%%%%%%%%%%%%%%%%%%%%%%%%%%%%%%%%%%%%%%%%%%%%%%%%%%%%%%%%%%%%%%%%%%%%%%%%%%%%
\section{Model}
\label{sec:model}

Our thought ratchet-and-pawl mechanism, which is nothing but an analogue of a famous Feynman-Smoluchowski ratchet, is illustrated in Fig.~\ref{fig:model}. We assume that ratchet wheels are synchronized (represented by a ``stick'' connecting the wheels) and immersed in a fluid at the temperature $T_x$. The wheels rotate randomly due to collisions with molecules of the  surrounding fluid. Furthermore, a small pawl (the T-shaped part) is placed in between the wheels. In our model it is not connected to a spring and it jiggles randomly between the wheels in horizontal direction only. The random motion is caused by molecules of a fluid from the second reservoir (small rectangle) at the temperature $T_y$. 

We model stochastic dynamics of the mechanical device by an overdamped 
diffusion\footnote{Of course, one may use a more fundamental and less tractable underdamped description, which, however, is expected to yield qualitatively similar results in the long-time limit \cite{SekimotoBook, NakagawaKomatsuJPSJ2005, NakagawaKomatsu2005}.} 
of two coupled degrees of freedom denoted as $x(t)$ and $y(t)$. 
The first, $x(t)$, is the angle of rotation of the wheels [in units of $2\pi$, thus $x\in[0,1)$ within one period]. The second, $y(t)$, represents position of the pawl between the teeth. 
The motion of the pawl is restricted to a narrow region between ratchet teeth. We describe the mutual repulsive interaction between teeth and pawl by the potential
\begin{equation}
\label{potentialGEN} 
U(x,y)= \frac{k(x)}{ 2\varepsilon^{2} } y^{2}, \qquad k(x) = k(x+1).
\end{equation} 
For $y=0$ the pawl is exactly in the middle of the wheels, the interaction potential is zero and wheels can rotate freely. When $y>0$, the pawl is closer to the right wheel and interacts with it repulsively (similarly for $y<0$). 
The width of the region where the pawl can diffuse is controlled by the  $1$-periodic ``spring constant'' $k(x)$, the fraction $\varepsilon/\sqrt{k(x)}$ is associated with the distance between the teeth for a given $x$. 

The actual shape of the ratchet teeth is reflected in the functional form of $k(x)$. In principle three qualitatively different cases may occur. 
First, when $k(x)$ is an asymmetric function of $x$, such as the frequently used function \cite{Reimann}
\begin{equation} 
\label{SpringConst} 
k(x) = 2-\frac{\sin(2\pi x)}{2} -\frac{ \sin(4 \pi x)}{12}. 
\end{equation} 
This case corresponds to {\em asymmetric teeth} as those illustrated in Fig.~\ref{fig:model}. We will show below that for $T_x\neq T_y$ both the mean heat flow $\dot{Q}$ through the system and the mean velocity $v$ are nonzero in this case. Second,  $k(x)$ can be a symmetric function, like $k(x) = 2-\sin(2\pi x)/2$, which corresponds to {\em symmetric teeth}. In this second case the device is not able to work as a heat engine: the mean velocity $v$ vanishes and only heat flow can be nonzero. The third case occurs if $k(x)$ is a constant. The latter situation corresponds to the wheels with no teeth. There, the pawl decouples from the wheels and each degree of freedom equilibrates in its own heat bath. In the following we stick to the most interesting first case, where the device can act as a ratchet. For a further insight into a particular choice of the potential we refer to the last paragraph of Sec.~\ref{sec:conclusions}. 

Summing up the above description, we arrive at the Langevin equations for coordinates $x(t)$, $y(t)$,
\begin{equation} 
\label{Langevin}
\frac{{ d} x}{{ d}t} = -   \frac{k'(x) }{2 \varepsilon^{2}}y^{2}  + \sqrt{2 T_x }\, \xi_{x}(t), \qquad 
\frac{{ d} y}{{ d}t} = - \frac{k(x)}{\varepsilon^{2}} y + \sqrt{2 T_y }\, \xi_{y}(t),  
\end{equation}
where $\xi_{i}(t)$ stands for the delta-correlated Gaussian white noise with  $\left<\xi_{i}(t)\xi_{j}(t')\right> = \delta_{ij} \delta(t-t')$, and $\left<\xi_{i}(t)\right> =0$, $i,j=x,y$. Throughout the paper we assume that the Boltzmann's constant and the friction coefficients are equal to one. Numerical integration of these equations reveals the desired intriguing functionality of our device. When $T_x \neq T_y$,  the wheels indeed rotate on average in one direction. A few generated trajectories of $x(t)$ are plotted in Fig.~\ref{fig:model}. 

Equivalently, the Langevin equations~(\ref{Langevin}) describe two-dimensional motion of a {\em single} Brownian particle confined to a periodic channel. Potential within one period of the channel is plotted in the upper right panel of Fig.~\ref{fig:model}. The channel central line runs along $x$ direction at $y=0$.  Thus, in the following we refer to the motion along $x$ as to {\em longitudinal} motion. In the {\em transversal} direction $y$ the potential increases without bounds and thus restricts the particle to diffuse along $x$. Note that in contrast to {\em entropic} transport, our channel is ``soft'' since its walls are represented by the potential. 

%%%%%%%%%%%%%%%%%%%%%%%%%%%%%%%%%%%%%%%%%%%%%%%%%%%%%%%%%%%%%%%%%%%%%%%%%%%%%%%%%%%%%%%%%%%%%%%%%%%%%%%%%%%%%%%%%%%%%%%%%%%%%%%%%%%
%%%%%%%%%%%%%%%%%%%%%%%%%%%%%%%%%%%%%%%%%%%%%%%%%%%%%%%%%%%%%%%%%%%%%%%%%%%%%%%%%%%%%%%%%%%%%%%%%%%%%%%%%%%%%%%%%%%%%%%%%%%%%%%%%%%
\section{Solution of the Fokker-Planck equation}
\label{sec:FP}

Let us now summarize main ideas behind the perturbative small-width solution of the two-dimensional Fokker-Planck equation corresponding to Langevin equations~(\ref{Langevin}). Detailed exposition, including all steps of derivations, have been published in our recent mathematical work, Ref.~\cite{RatchetJSTAT2016}. In next sections we exploit this solution to investigate dynamics and energetics of the model.

For narrow channels ($\varepsilon\ll 1$) the transversal dynamics is much faster than the longitudinal one, because the ratio of the relaxation time for the transversal dynamics to the relaxation time for the longitudinal dynamics is of the order of $\varepsilon^{2}/k(x)$ and it decreases with decreasing channel width ($\varepsilon\to 0$). 
The rescaling 
\begin{equation} 
\zeta = \frac{y}{\varepsilon}
\end{equation}
of the coordinate $y$ helps to separate the fast transversal and the slow longitudinal dynamics in narrow channels.
The Fokker-Planck equation corresponding to Eqs.~(\ref{Langevin}) then reads 
\begin{equation} 
\label{FokkerPlanck}
\varepsilon^{2} \left( \frac{\partial p}{ \partial t}+ \frac{\partial {j}_{x}}{\partial x} \right)   
+\frac{\partial {j}_{\zeta}}{\partial \zeta} =0 ,  
\end{equation} 
with the longitudinal and transversal components of the probability current given by 
\begin{equation}
\fl 
 {j}_{x} =
-\left[ T_x \frac{\partial }{\partial x}  +   \frac{ k'(x) }{2 }\zeta^{2} \right]p(x,\zeta,t ) ,\qquad 
 \frac{1}{\varepsilon^2}{j}_{\zeta}  = - \frac{1}{\varepsilon^2}\left[
T_y \frac{\partial }{\partial \zeta}  +  k(x)\zeta \right] 
p(x,\zeta,t), 
\label{currents}
\end{equation} 
respectively. In the long-time limit (steady-state), it is convenient to work with so called {\em reduced}  probability density and current defined as 
\begin{equation}
\label{reducedSUM}
P(x,\zeta,t) = \sum_{m=-\infty}^{+\infty } p(x+m,\zeta,t),
\qquad 
\mathbf{J}(x,\zeta,t) = \sum_{m=-\infty}^{+\infty } \mathbf{j}(x+m,\zeta,t).
\end{equation} 
 The reduced current $\mathbf{J}$ has components $J_x$ and $J_\zeta$ given in 
 Eqs.~(\ref{currents})\footnote{Note that $J_\zeta$ is actually the transversal reduced probability current multiplied by $\varepsilon^2$.} but with the PDF  $p$ replaced by the reduced PDF $P$.  In contrast to $p$, the reduced PDF  is  periodic with the period of the potential and it is normalized to unity in the potential unit cell,
\begin{equation} 
\label{periodicityANDnormalization}
P(x+1,\zeta)=P(x,\zeta),\qquad
\int_{0}^{1}d x  \int_{-\infty}^{+\infty}d \zeta\,  P(x,\zeta) = 1. 
\end{equation} 
See the review \cite{Reimann} for more details.

 In the long-time limit, the reduced PDF approaches its stationary form, which solves the stationary  Fokker-Planck equation, i.e., Eq.~(\ref{FokkerPlanck}) with $\partial P/\partial t =0$, subject to the normalization and periodicity conditions~(\ref{periodicityANDnormalization}). For a narrow channel, the stationary PDF and current can be represented by series in $\varepsilon^{2}$, 
\begin{equation} 
\label{expansion}
P  = P^{(0)} + \varepsilon^{2} P^{(1)} + \ldots, \qquad  
{\mathbf J} = {\mathbf J}^{(0)} + \varepsilon^{2} {\mathbf J}^{(1)} + \dots 
\end{equation} 
where individual components of the current $\mathbf{J}^{({n})}$ are defined as in Eqs.~(\ref{currents}) with  corresponding $P^{(n)}$ instead of $p$. 
Inserting these expansions into the stationary ($\partial P/\partial t =0$) Fokker-Planck equation yields
\begin{equation} 
\label{StacJeqs} 
\frac{\partial J^{(0)}_{\zeta}}{\partial \zeta} =0, \qquad   
 \frac{\partial J^{(n)}_{x}}{\partial x} + \frac{\partial J^{(n+1)}_{\zeta}}{\partial \zeta} =0,
 \qquad n=0,1,2,\ldots 
\end{equation}
Eqs.~(\ref{StacJeqs}) give us differential equation for any $P^{(n)}$ in terms of $P^{(n-1)}$ and thus they can in principle be solved recursively for any $n$ \cite{RatchetJSTAT2016}.

The principal part of $P$ determining its global shape is given by $P^{(0)}$. It follows from Eqs.~(\ref{StacJeqs}) that 
\begin{equation}
\label{P0A}
P^{(0)}(x,\zeta) =\mathcal{N} \left( {\frac{2\pi T_y }{k(x )}}
 \right)^{\frac{T_y-T_x}{2T_x}} \exp\! \left({-\frac{k(x)}{2 T_y} \zeta^{2} }\right),
\end{equation} 
where we choose $\mathcal{N}$ such that $P^{(0)}$ is normalized to one in the unit cell. Similarity of $P^{(0)}$ with the Gibbs canonical distribution is not accidental.In the narrow channel, the Gibbs equilibrium with the transversal heat bath holds locally for any $x$, which is a consequence of the separation of fast {transversal} and slow {longitudinal} degrees of freedom embodied in hierarchy~(\ref{StacJeqs}). 
The local width of the channel enters both the exponent and the $x$-dependent pre-exponential factor. Note that any PDF in the form $A(x)\exp\! \left({-{k(x)} \zeta^{2}/{2 T_y} }\right)$ satisfies the first of Eqs.~(\ref{StacJeqs}). The  pre-exponential factor $A(x)$ is  obtained from a second-order ordinary differential equation, which follows from the second of Eqs.~(\ref{StacJeqs}) for $n=0$ after integration with respect to $\zeta$ \cite{RatchetJSTAT2016}. These two steps (first guessing the $\zeta$-dependence, then deriving $x$-dependent terms) should be repeated when solving the hierarchy~(\ref{StacJeqs}) also for higher $n$.  

The hard task solved in Ref.~\cite{RatchetJSTAT2016} was to get $P^{(1)}$. This small correction  is crucial for capturing the ratchet effect, absent in $P^{(0)}$, because the local-equilibrium form of $P^{(0)}$ cannot support any global current through the system. 
Fortunately, it is possible to exploit the symmetry of the parabolic potential to get the exact expression for $P^{(1)}$ as the sum of three terms,  
\begin{equation} 
\label{P1}
P^{(1)}(x,\zeta ) = \sum_{n=0}^{2}C_{n}(x) \zeta^{2 n} 
{\rm e}^{-\frac{k(x)}{2 T_y} \zeta^{2} }, 
\end{equation} 
with coefficients given by 
\begin{eqnarray}
C_{1}(x) & = & - \frac{T_x}{2 T_y}  
\left( \frac{\partial^{2} P^{(0)}}{\partial x^{2} } \right)_{\zeta=0},
\qquad
C_{2}(x) =  \frac{T_x}{8T_y^{2}}\frac{{\rm d} k}{{\rm d}x }
\left( \frac{\partial P^{(0)}}{\partial x } \right)_{\zeta=0}, 
\\
\label{C0result}
C_{0}(x) & = & \left[ M_0(x) \right]^{\frac{T_y-T_x}{T_x}}\left[ \mathcal{M} + 
\frac{1}{T_x} 
\int_{0}^{x} {\rm d}x' \, 
\frac{R(x')-v_{1}}{\left[ M_0(x') \right]^{T_y/T_x}} 
\right],
\end{eqnarray}
where the last one, is expressed using auxiliary periodic functions 
\begin{equation}
\fl
R(x) =
T_y \left[C_{1}(x) \frac{{\rm d} M_{1}}{{\rm d}x} + 
C_{2}(x)\frac{{\rm d} M_{2}}{{\rm d}x} \right] 
 - T_x \frac{\rm d}{{\rm d}x} \left[C_{1}(x)M_{1}(x)+C_{2}(x)M_{2}(x) \right]  ,
\end{equation}
\begin{equation}
\label{Mn}
M_{n}(x) = \int^{+\infty}_{-\infty}{\rm d}\zeta \, \zeta^{2 n}\, 
{\rm e}^{-\frac{k(x)}{2 T_y} \zeta^{2} } ,\qquad n=0,1,2.
\end{equation}

The two coefficients $C_1$, $C_2$ are simple functions of $x$ and temperatures. The coefficient $C_0$, however, looks quite elaborate. Luckily for us it does not appear in any expression in what follows. The only important part of $C_0$ is the integration constant $v_1$, which determines the mean velocity of the particle~(\ref{v1TxTy}). It follows from the requirement of periodicity: $C_0(x)=C_0(x+m)$ for any integer $m$. (The second integration constant, $\mathcal{M}$, should be chosen in accordance with the normalization condition~(\ref{periodicityANDnormalization}) such that
$\int_{0}^{1}d x \int_{-\infty}^{+\infty}d \zeta  P^{(1)}(x,\zeta) = 0$.)

The expansion $P\approx P^{(0)}+\varepsilon^{2} P^{(1)}$, is rather convenient. It yields a simple closed analytical solution for any non-linear function $k(x)$ and, more importantly, {\em it is not restricted to the small temperature difference} allowing us to explore inherently far-from-equilibrium phenomena. There are, however, two practical limits of validity which we now emphasize. 
First, the expansion~(\ref{StacJeqs}) holds uniformly in the channel provided the inequality 
$ \varepsilon^{2} / k(x) \ll 1$, is fulfilled for all $x$. 
Otherwise, the expansion could fail locally in extremely wide regions, where $k(x) \approx \varepsilon$ or $ k(x) \approx \varepsilon^{2}$. The function $k(x)$ in Eq.~(\ref{SpringConst}), chosen for graphical illustrations, satisfies the above inequality. 
Second, high longitudinal temperature $T_x$ ruins precision of the approximation, which we also demonstrate in the following. From Eqs.~(\ref{FokkerPlanck}) and (\ref{currents}) we observe that for large $T_x$, the product $\varepsilon^{2}T_x$ appearing in the $\varepsilon^{2}j_x$ term, is no-longer small and different approximation scheme should be used.  
These two limitations are specific for the case of a narrow channel with {\em soft} walls. They do not arise for hard-wall channels, where the expansion, similar in spirit to the present one, was used for the first time \cite{Laachi2007}.

%%%%%%%%%%%%%%%%%%%%%%%%%%%%%%%%%%%%%%%%%%%%%%%%%%%%%%%%%%%%%%%%%%%%%%%%%%%%%%%%%%%%%%%%%%%%%%%%%%%%%%%%%%%%%%%%%%%%%
%%%%%%%%%%%%%%%%%%%%%%%%%%%%%%%%%%%%%%%%%%%%%%%%%%%%%%%%%%%%%%%%%%%%%%%%%%%%%%%%%%%%%%%%%%%%%%%%%%%%%%%%%%%%%%%%%%%%%
%%%%%%%%%%%%%%%%%%%%%%%%%%%%%%%%%%%%%%%%%%%%%%%%%%%%%%%%%%%%%%%%%%%%%%%%%%%%%%%%%%%%%%%%%%%%%%%%%%%%%%%%%%%%%%%%%%%%%
\section{Mean velocity and heat current} 
\label{sec:vQ}

Basic quantities which characterize ratchet performance are the mean rotation velocity of the wheels $v(T_x,T_y)$, and the heat flow between reservoirs $\dot{Q}(T_x, T_y)$. They are defined as long-time (steady-state) averages given by 
\begin{equation} 
\label{vQ_deff}
v(T_x, T_y) = \lim_{t\to \infty} \frac{\left< x(t) \right> }{t},
\qquad 
\dot{Q}(T_x, T_y) = \lim_{t\to \infty} \frac{ Q_{y}(t) }{t},
\end{equation} 
where $\left< x(t) \right>$ is average particle position and $Q_{y}(t)$ denotes total mean amount of heat \emph{accepted} by the transversal heat bath  during the time interval $(0,t)$. We adopt this definition of heat flow (positive when heat flows from $x$ to $y$ reservoir) because it has the same sign as the mean velocity. Indeed, for $T_x<T_y$ (heat flow from transversal to longitudinal) the mean velocity is negative, as one can infer already from Fig.~\ref{fig:model}. 

The both quantities follow directly from results of the preceding section, namely, from calculated components of the stationary reduced probability current   ($J_x$, $J_\zeta$). The mean velocity is just integral of $J_x$ over the unit cell of the 
potential,\footnote{\label{footnote3} $\left<dx/dt\right>=\int dx \int dy \left<dx/dt  \delta(x-x(t))\delta(y-y(t))\right>=\int dx \int d\zeta J_x(x,\zeta,t)$, which in the long-time limit yields Eq.~(\ref{velocitydef}), see Ref.\ \cite{Reimann}}
\begin{equation}
\label{velocitydef} 
 v(T_x,T_y)  = 
\int_{0}^{1}d x  \int_{-\infty}^{+\infty}d \zeta\,  J_{x}(x,\zeta),
\end{equation} 
the average heat flow can be obtained from $J_\zeta$,
\begin{equation}
\label{heatflowdef} 
\dot{Q}(T_x,T_y)=  - \frac{1}{\varepsilon^2}\int_{0}^{1}dx \int_{-\infty}^{+\infty}d\zeta\,  k(x) \zeta J_\zeta(x,\zeta).
\end{equation}
In the last equation we have used definition of heat standard in stochastic thermodynamics \cite{Sekimoto1997, SekimotoBook, Seifert2012}:
According to the first law of thermodynamics and Eq.~(\ref{FokkerPlanck}), the internal energy of the system, $E = \int_0^1 dx \int_
{-\infty}^{\infty} dy U(x,y) P(x,y)$, changes in course of time as 
\begin{equation}
\fl
\label{firstlaw}
\dot{E} =\! -(\dot{Q}_x + \dot{Q}_y) =\!\! \int_0^1 \!\!\! dx \int_
{-\infty}^{\infty}\!\! d\zeta \frac{\partial U(x,\varepsilon\zeta)}{\partial x} J_x(x,\zeta) + \frac{1}{\varepsilon^2}\int_0^1\!\!\! dx \int_
{-\infty}^{\infty}\!\! d\zeta \frac{\partial U(x,\varepsilon\zeta)}{\partial \zeta} J_\zeta(x,\zeta),
\end{equation}
where $\dot{Q}_x$ is the mean heat flow into the $x$ reservoir. 
According to the definition~(\ref{vQ_deff}), in the steady state we have $\dot{Q}_y=\dot{Q}$ and the expression~(\ref{heatflowdef}) follows directly from the last term on the right-hand side of Eq.~(\ref{firstlaw}).

Our convention used in definition of heat is that heat is positive when it flows from the system to a heat bath. Thus $\dot{Q}_x$ ($\dot{Q}_y$) denotes the average heat flow {\em accepted} by the longitudinal (transversal) bath. According to the conservation of energy (\ref{firstlaw}) the following relation between energy flows holds: $\dot{E}=-(\dot{Q}_x+\dot{Q}_y$). In the steady state, the mean energy of the system is constant, hence we have $\dot{E}=0$ and $\dot{Q}_x=-\dot{Q}_y$. On the other hand, in the transient regime (before the steady state is established), the system energy may change and in general $\dot{E} \neq 0$. The expression of the steady-state heat flow into the transversal bath~(\ref{heatflowdef}), $\dot{Q}=\displaystyle \lim_{t\to\infty}\dot{Q}_y$ in terms of the integrated probability current can be also justified as follows. The heat flow $\dot{Q}_y$ is identified from the first law of thermodynamics \cite{Sekimoto1997, SekimotoBook, Seifert2012}, 
$\dot{E}= \left< d U(x(t),y(t))/dt \right> 
= \left< \dot{x} (\partial U/\partial x) \right> + 
 \left< \dot{y} (\partial U/\partial y) \right>$,
as the second term, which corresponds to the change of potential energy, when the transversal coordinate is changed,
$\dot{Q}_y =- \left< \dot{y} (\partial U/\partial y) \right> $. Using formal manipulation from the footnote~\ref{footnote3}, we arrive directly at Eq.~(\ref{heatflowdef}).

Expansion in the channel width, developed in the last section, yields the stationary probability current in the form $\mathbf{J}\approx \mathbf{J}^{(0)}+\varepsilon^{2} \mathbf{J}^{(1)}$. Hence Eqs.~(\ref{velocitydef}), (\ref{heatflowdef}) become  
\begin{equation}
\label{vQ01}
\fl
v(T_x, T_y) \approx v_{0}(T_{x},T_{y} )+ \varepsilon^{2} v_{1}(T_{x},T_{y} ), \qquad 
\dot{Q}(T_x, T_y) \approx \dot{Q}_{0}(T_{x},T_{y} )+ \varepsilon^{2} \dot{Q}_{1}(T_{x},T_{y} ),
\end{equation}
respectively. The individual summands follow from the corresponding parts of the current, e.g., 
$ v_{n}(T_x, T_y ) = \int_{0}^{1}d x  \int_{-\infty}^{+\infty}d \zeta\,  J_{x}^{(n)}(x,\zeta)$, $n=0,1$, 
and similarly for $\dot{Q}_{n}(T_x, T_y )$. 

Frequently, all important physical effects  are (at least qualitatively) contained in the lowest order of the Fick-Jacobs theory, described by the simple PDF $P^{(0)}$, Eq.~(\ref{P0A}). In our case, however, the fundamental assumption of the Fick-Jacobs approximation renders the lowest order useless for explanation of the ratchet effect. 
The local equilibrium with the transversal heat bath implies that there is no net heat transfer into the transversal heat bath, therefore also the heat flow to the longitudinal bath is on average zero. And when there is no heat flow,  the system cannot act as the heat engine. In Eqs.~(\ref{vQ01}) we therefore have $v_{0}(T_{x},T_{y} )=0$, and $\dot{Q}_{0}(T_{x},T_{y} )=0$, for any $T_x$, $T_y$. The ratchet effect is covered by the correction $P^{(1)}$,  Eq.~(\ref{P1}), which disrupts the local equilibrium, and hence we have
\begin{equation} 
\label{vQexpanded} 
v(T_x, T_y) \approx  \varepsilon^{2} v_{1}(T_{x},T_{y} ) , \qquad
\dot{Q}(T_x, T_y) \approx \varepsilon^{2} \dot{Q}_{1}(T_{x},T_{y}).
\end{equation}

The both quantities can be given by simple expressions. The mean velocity follows from the requirement that the coefficient $C_0(x)$, Eq.~(\ref{C0result}), is a $1$-periodic function of $x$. We get
\begin{equation}
\label{v1TxTy}
v_1(T_{x},T_{y} ) = \frac{
\int_{0}^{1} {\rm d}x \, 
{R(x) \left[ M_0(x) \right]^{-T_y/T_x}}
 }{ \int_{0}^{1} {\rm d}x \, 
{\left[ M_0(x) \right]^{-T_y/T_x}} }.
\end{equation}
The mean heat flow can be evaluated directly from its definition~(\ref{heatflowdef}) inserting there $J_\zeta^{(1)}$ as computed using~(\ref{P1}). In the course of the derivation one finds that the zeroth term of the sum (\ref{P1}) representing $P^{(1)}$ makes no contribution to the current $J_\zeta^{(1)}$. Simplification of the remaining terms gives us
\begin{equation}
\label{Qexplicit}
\dot{Q}_1(T_x,T_y) =  \frac{T_x T_y}{\varepsilon^2} \int_0^{1}dx\,M_0(x) \frac{k'(x)}{k(x)} \left( \frac{\partial P^{(0)}}{\partial x } \right)_{\zeta=0}.
\end{equation}

The both main results of the present section, Eqs.~(\ref{v1TxTy}) and (\ref{Qexplicit}), can be further recast into simple scaling forms which reveals their temperature dependence. Note that temperatures $T_x$, $T_y$ occur in all auxiliary functions used in Eqs.~(\ref{v1TxTy}) and (\ref{Qexplicit}). A closer look reveals that $T_x$ enters all expressions only in the combination $T_x/T_y$ and, at the same time almost all powers of $T_y$ cancel. Eventually, we end up with expressions 
\begin{equation}
v(T_x,T_y) \approx \varepsilon^{2}T_y^{2}\, \mathcal{V}(T_x/T_y) , \qquad
\dot{Q}(T_x,T_y) \approx T_y^{2}\, \mathcal{Q}(T_x/T_y), 
\label{eq:V_and_Q_final}
\end{equation}
where the {\em master functions} $\mathcal{V}(T_x/T_y) = v_1(T_x,T_y)/T_y^2$ and $\mathcal{Q}(T_x/T_y) = \varepsilon^2 \dot{Q}_1(T_x,T_y)/T_y^2$ depend on the combination $T_x/T_y$ only. Hence they characterize the ratchet performance in a universal manner depending only on the ratio of $T_x$ and $T_y$ and not on individual values of temperatures.  
More importantly, the master functions yield  a {\em figure of merit} of the ratchet, $\eta$,  
\begin{equation}
\eta(T_x/T_y) = \frac{v}{\dot{Q}}\approx \varepsilon^2 \frac{\mathcal{V}(T_x /T_y)}{\mathcal{Q}(T_x/T_y)},
\label{eq:eta}
\end{equation} 
which in the leading order in $\varepsilon$ depends only on the ratio of temperatures $T_x/T_y$. The figure of merit $\eta$ quantifies how much heat must flow through the system in order to maintain the rotation velocity $v$. It is large when small heat current (small dissipation) is accompanied by large velocity and vice versa.
The figure of merit $\eta$ is different from the standard efficiency (output power/input heat flow) used to characterize steady-state heat engines \cite{BroeckPRL2005, RyabovHolubec2016, SekimotoBook, Seifert2012}. This standard ``energetic'' efficiency is bounded by unity and reflects effectiveness of energy transformation from accepted heat to a useful work output. The figure of merit $\eta$ describes rather a ``kinetic performance'' of the model and it is in principle not bounded from above. In order to define the standard efficiency in our model, one should introduce an external load, against which the ratchet would perform work. This extension of the model is currently under investigation.

Equations (\ref{eq:V_and_Q_final})  and (\ref{eq:eta}) supplemented by exact expressions (\ref{v1TxTy}) and (\ref{Qexplicit}) reveal somewhat unintuitive behavior of the velocity, the heat current and the ratchet figure of merit with respect to the channel width. While both the velocity and the figure of merit are proportional to $\varepsilon^2$, the heat flow becomes nonzero and independent of $\varepsilon$ as $\varepsilon\to 0$. 
This means that $\dot{Q}\neq 0$ for arbitrarily narrow channel whenever $T_x\neq T_y$ and $\varepsilon$ is arbitrary small but nonzero. On the other hand, for $\varepsilon = 0$ the transversal and longitudinal degrees of freedom decouple and thus $\dot{Q}=0$. The heat flow thus experiences a discontinuity when $\varepsilon$ changes from arbitrarily small positive value to zero. This behavior demonstrates what we knew from the very beginning: 
the system with small, but positive $\varepsilon$ qualitatively differs from the system with $\varepsilon = 0$. While the former case represents diffusion in a two-dimensional energy landscape, the latter  stands just for a Brownian motion on a line. Similar reasoning explains also another difference between the velocity and the heat flow: the velocity vanishes for symmetric potentials, i.e., for potentials with symmetric $k(x)$ corresponding to symmetric ratchet teeth, while the heat flow is non-zero whenever $k(x)$ varies with $x$ and $T_x\neq T_y$. 
The nonzero velocity is achievable only in cases where the ratchet teeth are asymmetry, but the heat flows between the reservoirs whenever they are coupled by a nontrivial interaction between wheels and the pawl.

%%%%%%%%%%%%%%%%%%%%%%%%%%%%%%%%%%%%%%%%%%%%%%%%%%%%%%%%%%%%%%%%%%%%%%%%%%%%%%%%%%%%%%%%%%%%%%%%%%%
%%%%%%%%%%%%%%%%%%%%%%%%%%%%%%%%%%%%%%%%%%%%%%%%%%%%%%%%%%%%%%%%%%%%%%%%%%%%%%%%%%%%%%%%%%%%%%%%%%%
\begin{figure}[t!]
\begin{center}
\includegraphics[width=.95\linewidth]{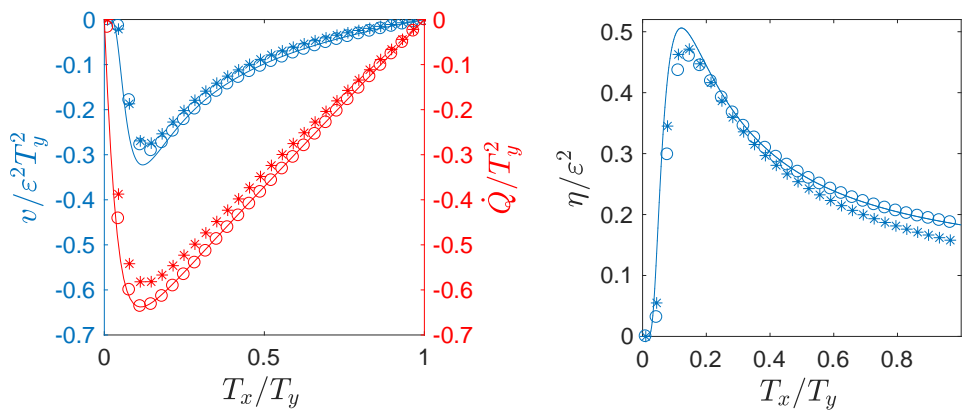}
\caption{The mean velocity $v$, the heat flow $\dot{Q}$ (left) and the figure of merit $\eta$ (right) as the functions of reservoir temperatures for $T_x<T_y$. Approximate analytical curves (solid lines) are plotted using Eqs.~(\ref{v1TxTy}) and (\ref{Qexplicit}). In the numerics we set $\varepsilon=0.01$; $T_y = 1$ ($\circ$) and $T_y = 100$ ($\ast$).}
\label{fig:TxTy}
\end{center}
\end{figure} 
%%%%%%%%%%%%%%%%%%%%%%%%%%%%%%%%%%%%%%%%%%%%%%%%%%%%%%%%%%%%%%%%%%%%%%%%%%%%%%%%%%%%%%%%%%%%%%%%%%%
%%%%%%%%%%%%%%%%%%%%%%%%%%%%%%%%%%%%%%%%%%%%%%%%%%%%%%%%%%%%%%%%%%%%%%%%%%%%%%%%%%%%%%%%%%%%%%%%%%% 

\begin{figure}[t!]
\begin{center}
\includegraphics[width=.95\linewidth]{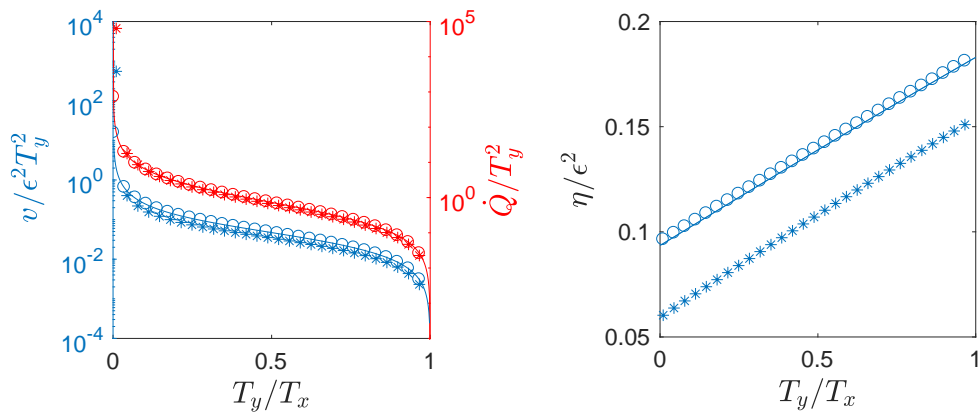}
\caption{The mean velocity $v$, the heat flow $\dot{Q}$ (left) and the figure of merit $\eta$ (right) as the functions of reservoir temperatures for $T_x>T_y$. Approximate analytical curves (solid lines) are plotted using Eqs.~(\ref{eq:V_and_Q_final}) and (\ref{eq:eta}). In the numerics we set $\varepsilon=0.01$; $T_x = 1$ ($\circ$) and $T_x = 100$ ($\ast$).}
\label{fig:TyTx}
\end{center}
\end{figure}

The  master functions $\mathcal{V}(T_x/T_y)$ and $\mathcal{Q}(T_x/T_y)$ are plotted in left panels and the figure of merit $\eta$ in right panels of Figs.~\ref{fig:TxTy} and \ref{fig:TyTx} together with the corresponding quantities obtained numerically by discretizing the underlying Fokker-Planck equation and numerically finding the steady state of the discrete model (see appendix in Ref.~\cite{RatchetJSTAT2016} for detailed description of the numerics). In Fig.~\ref{fig:TxTy} the transversal temperature $T_y$ is fixed and the longitudinal temperature $T_x$ is varied from $0$ to $T_y$. We see that the agreement between approximate analytical curves and the numerical data is very good for the data obtained for $T_y = 1$ (circles), while the agreement is only qualitative for the data calculated using $T_y = 100$ (stars). This is because in the first case $T_x$ is always small and the assumptions used in the analytical derivation are valid, while in the later case $T_x$ is relatively large and $\varepsilon^2 T_x$ is not small enough (see the last paragraph of Sec.~\ref{sec:FP}). For $T_x<T_y$, the particle moves on average to the left ($v<0$) and the heat flows from the transversal (hot) to the longitudinal (cold) reservoir ($\dot{Q} < 0$). Both these quantities vanish for $T_x/T_y = 1$ and for $T_x = 0$. For $T_x = T_y$ the ratchet attains thermal equilibrium where all flows are zero, for $T_x = 0$ the longitudinal thermal noise is switched off and the particle feels in the $x$ direction just the deterministic force with no global bias. The fact that between these two points $v < 0$ and $\dot{Q} < 0$ immediately implies that the both functions attain a global minimum for some $T_x \in (0, T_y)$. 

The ratchet figure of merit $\eta$ in the right panel of Fig.~\ref{fig:TxTy} vanishes for $T_x \to 0$ (for $T_x \to 0$, $v$ converges to zero faster than $\dot{Q}$), reaches a constant value for $T_x \to T_y$ ($v$ and $\dot{Q}$ converges to zero at the same rate with $T_x \to T_y$) and attains a maximum value in between. This complicated behavior of $\eta$ clearly shows that the quantities $v$ and $\dot{Q}$ are not simply proportional to each other. Notice that in this regime the ratchet effect is readily visible on the level of individual trajectories shown in Fig.~\ref{fig:model}, in contrast to the case $T_x>T_y$ discussed next.

%%%%%%%%%%%%%%%%%%%%%%%%%%%%%%%%%%%%%%%%%%%%%%%%%%%%%%%%%%%%%%%%%%%%%%%%%%%%%%%%%%%%%%%%%%%%%%%%%%%
%%%%%%%%%%%%%%%%%%%%%%%%%%%%%%%%%%%%%%%%%%%%%%%%%%%%%%%%%%%%%%%%%%%%%%%%%%%%%%%%%%%%%%%%%%%%%%%%%%%
%%%%%%%%%%%%%%%%%%%%%%%%%%%%%%%%%%%%%%%%%%%%%%%%%%%%%%%%%%%%%%%%%%%%%%%%%%%%%%%%%%%%%%%%%%%%%%%%%%%
%%%%%%%%%%%%%%%%%%%%%%%%%%%%%%%%%%%%%%%%%%%%%%%%%%%%%%%%%%%%%%%%%%%%%%%%%%%%%%%%%%%%%%%%%%%%%%%%%%% 

Fig.~\ref{fig:TyTx} illustrates the ratchet performance in the regime $T_x>T_y$. In the numerics we have fixed the longitudinal temperature $T_x$ and varied the transversal temperature $T_y$ from $0$ to $T_x$. Again, analytical curves agree with the numerical data much better for a moderate longitudinal temperature than for large $T_x$, although the qualitative agreement is in both cases excellent (see in particular the right panel). In this regime, which was rather noisy on the level of individual trajectories (lower left panel in Fig.~\ref{fig:model}), the particle moves on average to the right ($v > 0$) and the heat flows to the transversal bath ($\dot{Q} > 0$). 
Also in this case $v$ and $\dot{Q}$ exhibit an extreme, because $v = \dot{Q} = 0$ for $T_x=T_y$, and also the both quantities vanish when $T_y \to 0$. In the latter limit the bath does not contain enough energy to push the particle away from the channel central at $y=0$, where $U=0$. The limit correspond to the cold pawl standing still just in between the wheels. The fact that in between $T_y = 0$ and $T_y = T_x$ we have $v>0$ and $\dot{Q} > 0$ then implies that the both variables exhibit a maximum for some $T_y \in (0, T_x)$. The maximum magnitude of the mean velocity obtainable in this regime ($T_x>T_y$) is by several orders of magnitude smaller than that in the regime $T_x<T_y$. This is why we cannot recognize any systematic drift when looking on corresponding trajectories in Fig.~\ref{fig:model}. 

If we compare the ratchet performance for the two regimes of operation ($T_x<T_y$ vs.\ $T_x>T_y$), on the level of individual trajectories we find that the regime $T_x<T_y$ is much more advantageous: it both gives a larger mean velocity of the particles and is not that noisy (see lower panels in Fig.~\ref{fig:model}). The fluctuations (quantified by an effective diffusion coefficient) have been studied in Ref.~\cite{RatchetJSTAT2016}, where it was shown that they are indeed much larger in the regime $T_x>T_y$. The mean particle velocity can be read from Figs.~\ref{fig:TxTy} and \ref{fig:TyTx}. In particular, in Fig.~\ref{fig:model} we took $T_x=1$, and $T_y=10$ for the regime $T_x<T_y$ and $T_x=10$, and $T_y=1$ for the regime $T_x>T_y$ and in the both cases $\epsilon^2=0.01$. In the regime $T_x<T_y$ we can read from Fig.~\ref{fig:TxTy} that for $T_x/T_y = 0.1$ the rescaled velocity $v/\epsilon^2T_y^2$ approximately equals to $-0.2$ and because we have $\epsilon^2 T_y^2 = 1$ for this regime, we estimate the mean velocity $v \approx -0.2$. On the other hand, in the regime $T_x>T_y$ we find from Fig.~\ref{fig:TyTx} that the rescaled velocity $v/\epsilon^2T_y^2$ for $T_x/T_y = 0.1$ is approximately $0.2$ and thus the average velocity is $v \approx 0.2 \epsilon^2T_y^2 = 0.002$ (quantitative disagreement with Fig.~\ref{fig:model} is due to the fact that $\varepsilon$ is not small enough there). Thus the mean velocity  in the regime $T_x>T_y$ is by two orders of magnitude smaller than that in the regime $T_x<T_y$.

The poor performance of the ratchet in the $T_x>T_y$ regime is evident also if we compare energetic costs per velocity, i.e., the figures of merit $\eta$ plotted in the right panel of Fig.~\ref{fig:TyTx}, against that in $T_x<T_y$ case shown in the right panel of Fig.~\ref{fig:TxTy}. The both figures of merit are equal for $T_y=T_x$. From Fig.~\ref{fig:TyTx} we see that $\eta$ in $T_x>T_y$ regime {\em decreases} with increasing temperature difference. Hence the maximum attained by $\eta$ in $T_y>T_x$ regime is the global maximum of the figure of merit. The only situation when the ratchet with $T_x>T_y$ can have a larger figure of merit is in an extreme situation when the transversal reservoirs (the pawl) is kept at a very cold temperature.

%%%%%%%%%%%%%%%%%%%%%%%%%%%%%%%%%%%%%%%%%%%%%%%%%%%%%%%%%%%%%%%%%%%%%%%%%%%%%%%%%%%%%%%%%%%%%%%%%%%%%%%%%%%%%%%%%%%%%%%%%%%%%%%%%%%
%%%%%%%%%%%%%%%%%%%%%%%%%%%%%%%%%%%%%%%%%%%%%%%%%%%%%%%%%%%%%%%%%%%%%%%%%%%%%%%%%%%%%%%%%%%%%%%%%%%%%%%%%%%%%%%%%%%%%%%%%%%%%%%%%%%
%%%%%%%%%%%%%%%%%%%%%%%%%%%%%%%%%%%%%%%%%%%%%%%%%%%%%%%%%%%%%%%%%%%%%%%%%%%%%%%%%%%%%%%%%%%%%%%%%%%%%%%%%%%%%%%%%%%%%%%%%%%%%%%%%%%
%%%%%%%%%%%%%%%%%%%%%%%%%%%%%%%%%%%%%%%%%%%%%%%%%%%%%%%%%%%%%%%%%%%%%%%%%%%%%%%%%%%%%%%%%%%%%%%%%%%%%%%%%%%%%%%%%%%%%%%%%%%%%%%%%%%

\section{Current circulation and local heat transfer}
\label{sec:circulation}

\begin{figure}[t!]
\begin{center}
\includegraphics[width=1\linewidth]{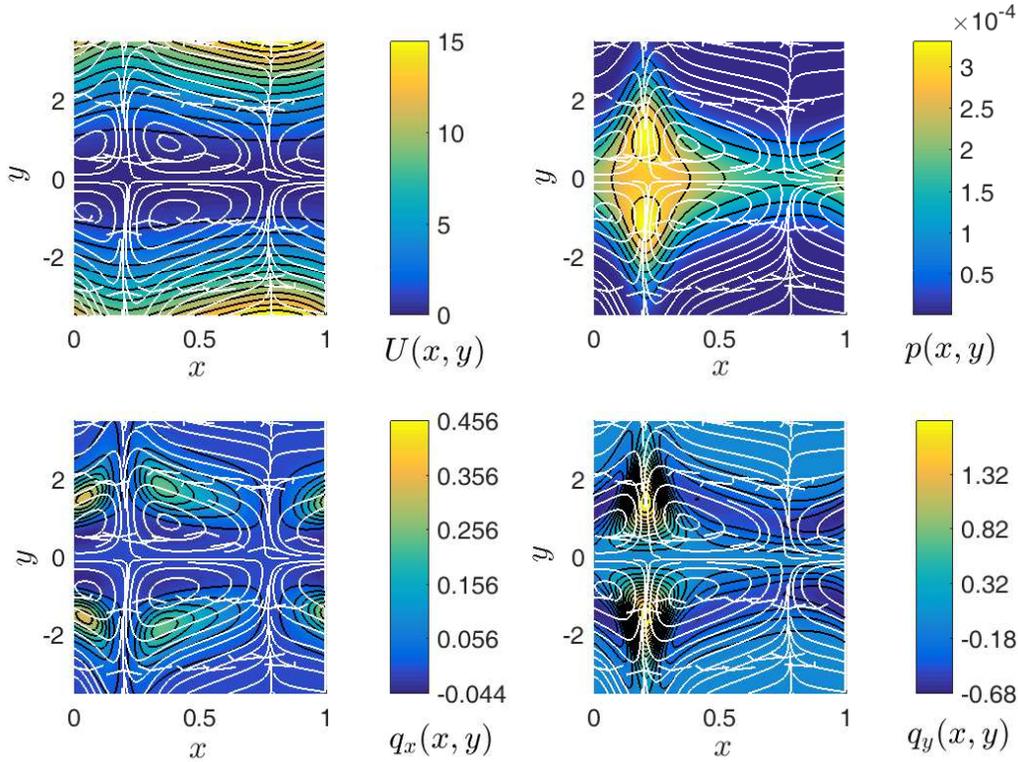}
\caption{Circulation of the probability current $\mathbf{j}(x,y)$ in a unit cell of the potential for $T_x < T_y$. The particle moves on average to the left ($v < 0$, see Fig.~\ref{fig:TxTy}). The current is plotted on top of the potential energy landscape (upper left), reduced probability density (upper right), local heat flow to the longitudinal reservoir (lower left) and local heat flow to the transversal reservoir (lower right). The data were obtained numerically using $T_x = 0.2$, $T_y = 2$ and $\varepsilon^2 = 1$.}
\label{fig:JTxmTy}
\end{center}
\end{figure}

\begin{figure}[t!]
\begin{center}
\includegraphics[width=1\linewidth]{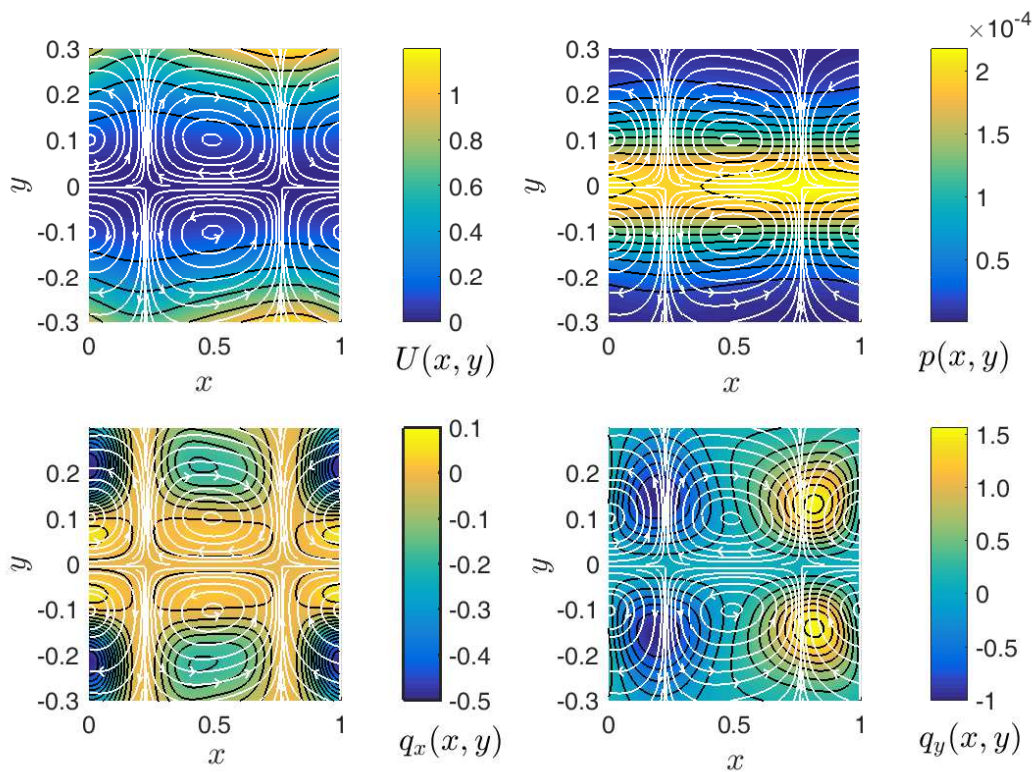}
\caption{Circulation of the probability current $\mathbf{j}(x,y)$ in a unit cell of the potential for $T_x > T_y$. The particle moves on average to the right ($v > 0$, see Fig.~\ref{fig:TyTx}). The current is plotted on top of the potential energy landscape (upper left), reduced probability density (upper right), local heat flow to the longitudinal reservoir (lower left) and local heat flow to the transversal reservoir (lower right). The data were obtained numerically using $T_x = 1$, $T_y = 0.2$ and $\varepsilon^2 = 0.1$.}
\label{fig:JTxvTy}
\end{center}
\end{figure}

As first noted in Ref.~\cite{Magnasco1998}, the origin of the ratchet effect is related to the circulation of probability current. Let us now illustrate this circulation and investigate its connection to the heat flow. 
To do this, we return  back to the physical coordinates $x$ and $y$. The reduced probability density for particle position in a unit cell of the potential, the longitudinal probability current and the transversal probability current in these coordinates can be calculated from the corresponding variables defined in the preceding sections as $p(x,y) = P(x,y/\varepsilon)/\varepsilon$, $j_x(x,y) = J_x(x,y/\varepsilon)/\varepsilon$ and $j_y(x,y) = J_\zeta(x,y/\varepsilon)/\varepsilon^2$, respectively. The streamlines of the vector $\mathbf{j}(x,y) = (j_x(x,y),j_y(x,y))$ are shown in Fig.~\ref{fig:JTxmTy} for $T_{x}<T_{y}$ and in Fig.~\ref{fig:JTxvTy} for $T_{x}>T_{y}$. In the both figures, we plot the streamlines on top of four important quantities: the potential landscape $U(x,y)$, the reduced probability density $p(x,y)$, the local heat flow to the longitudinal reservoir $q_x(x,y) = - j_x(x,y) \partial U(x,y)/\partial x $ and the local heat flow to the transversal reservoir $q_y(x,y) = - j_y(x,y) \partial U(x,y)/\partial y$. The shown data were obtained numerically using the same method as in the preceding section.

In the panels with $\mathbf{j}(x,y)$ plotted on top of $p(x,y)$, it is clearly visible how the probability currents feed maxima of the reduced probability distributions. These maxima are larger than corresponding equilibrium values of $P$ at any of the two temperatures $T_x$ and $T_y$. On the other hand, the panels where the current is plotted on top of the potential and the two heat flows show us at which coordinates the heat is drawn from the $x$ reservoir (if $j_x(x,y)$ points uphill in the $x$ direction then $q_x(x,y) < 0$) and from the $y$ reservoir (similarly, if $j_y(x,y)$ points uphill in the $y$ direction, $q_y(x,y)<0$). The panels with the heat flows demonstrate also how much heat is on average exchanged at a given point with the individual heat reservoirs.  Can the complex behavior depicted in Figs.~\ref{fig:JTxmTy} and \ref{fig:JTxvTy} be understood and predicted using simple physical arguments? In the rest of this section we will provide an affirmative answer.

%%%%%%%%%%%%%%%%%%%%%%%%%%%%%%%%%%%%%%%%%%%%%%%%%%%%%%%%%%%%%%%%%%%%%%%%%%%%%%%%%%%%%%%%%%%%%%%%%%%
%%%%%%%%%%%%%%%%%%%%%%%%%%%%%%%%%%%%%%%%%%%%%%%%%%%%%%%%%%%%%%%%%%%%%%%%%%%%%%%%%%%%%%%%%%%%%%%%%%%
%%%%%%%%%%%%%%%%%%%%%%%%%%%%%%%%%%%%%%%%%%%%%%%%%%%%%%%%%%%%%%%%%%%%%%%%%%%%%%%%%%%%%%%%%%%%%%%%%%%
%%%%%%%%%%%%%%%%%%%%%%%%%%%%%%%%%%%%%%%%%%%%%%%%%%%%%%%%%%%%%%%%%%%%%%%%%%%%%%%%%%%%%%%%%%%%%%%%%%% 

%%%%%%%%%%%%%%%%%%%%%%%%%%%%%%%%%%%%%%%%%%%%%%%%%%%%%%%%%%%%%%%%%%%%%%%%%%%%%%%%%%%%%%%%%%%%%%%%%%%
%%%%%%%%%%%%%%%%%%%%%%%%%%%%%%%%%%%%%%%%%%%%%%%%%%%%%%%%%%%%%%%%%%%%%%%%%%%%%%%%%%%%%%%%%%%%%%%%%%%
%%%%%%%%%%%%%%%%%%%%%%%%%%%%%%%%%%%%%%%%%%%%%%%%%%%%%%%%%%%%%%%%%%%%%%%%%%%%%%%%%%%%%%%%%%%%%%%%%%%
%%%%%%%%%%%%%%%%%%%%%%%%%%%%%%%%%%%%%%%%%%%%%%%%%%%%%%%%%%%%%%%%%%%%%%%%%%%%%%%%%%%%%%%%%%%%%%%%%%%

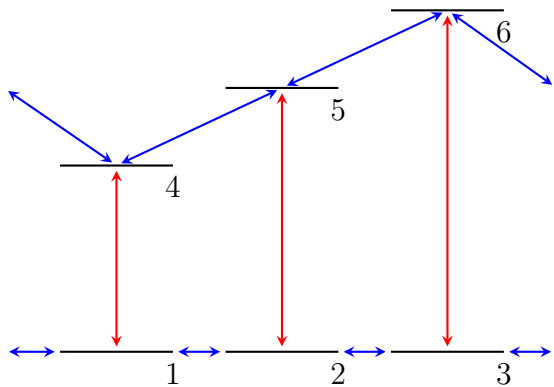
\begin{figure}
\centerline{
  % Resize it to 5cm wide.
  %\resizebox{7cm}{!}{
    \begin{tikzpicture}[
      scale=0.5,
      level/.style={thick,black},
      virtual/.style={thick,densely dashed},
      trans/.style={thick,<->,shorten >=2pt,shorten <=2pt,>=stealth},
      classical/.style={thin,double,<->,shorten >=4pt,shorten <=4pt,>=stealth}
    ]  
		% Draw lower energy levels.
		\draw[level] (3cm,-2em) node[right,below] {$1$} -- (0.0cm,-2em) node[right,below] {};
    \draw[level] (7.4cm,-2em) node[right,below] {$2$} -- (4.4cm,-2em) node[right,below] {};
    \draw[level] (11.8cm,-2em) node[right,below] {$3$} -- (8.8cm,-2em) node[right,below] {};
    % Draw upper energy levels.
		\draw[level] (3cm,+10em) node[right,below] {$4$} -- (0.0cm,+10.0em) node[right,below] {};
    \draw[level] (7.4cm,+15em) node[right,below] {$5$} -- (4.4cm,+15.0em) node[right,below] {};
    \draw[level] (11.8cm,+20em) node[right,below] {$6$} -- (8.8cm,+20em) node[right,below] {};
    % Draw lover x transitions.
		\draw[trans,blue] (-1.5cm,-2em) -- (0cm,-2em) node[midway,midway] {};
    \draw[trans,blue] (3cm,-2em) -- (4.4cm,-2em) node[midway,midway] {};
    \draw[trans,blue] (7.4cm,-2em) -- (8.8cm,-2em) node[midway,midway] {};
		\draw[trans,blue] (11.8cm,-2em) -- (13.2cm,-2em) node[midway,midway] {};
				% Draw upper x transitions.
		\draw[trans,blue] (-1.5cm,15em) -- (1.5cm,10em) node[midway,midway] {};
		\draw[trans,blue] (5.9cm,15em) -- (1.5cm,10em) node[midway,midway] {};
		\draw[trans,blue] (5.9cm,15em) -- (10.3cm,20em) node[midway,midway] {};
		\draw[trans,blue] (10.3cm,20em) -- (13.2cm,15em) node[midway,midway] {};
		% Draw upper y transitions.
		\draw[trans,red] (1.5cm,-2em) -- (1.5cm,10em) node[midway,midway] {};
		\draw[trans,red] (5.9cm,-2em) -- (5.9cm,15em) node[midway,midway] {};
    \draw[trans,red] (10.3cm,-2em) -- (10.3cm,20em) node[midway,midway] {};
    \end{tikzpicture}
  %}
}
\caption{Schematic illustration of the simple discrete model of the two-temperature ratchet. Red (blue) arrows depict transitions driven by the reservoir at $T_y$ ($T_x$).}
\label{fig:discrete_system}
\end{figure}

The derivation of the approximate formulas introduced in the preceding sections was based on the Fick-Jacobs theory developed for particles diffusing in a single thermal bath through asymmetric channels with hard walls. Although the expansion in the channel width works well both in the setup with hard walls and in our two-temperature soft-wall model, microscopic explanations of emerging ratchet effects differ. While the ratchet effects occurring in  hard-wall channels are of entropic origin \cite{BuradaPTRA2009, MartensPRE2011, Dagdug2012, RegueraPRL2012}, the ratchet effect for the present two-temperature soft-wall ratchet is of an energetic nature. In fact, the main operational principle of the present ratchet can be understood with the aid of a simple discrete ratchet model depicted in Fig.~\ref{fig:discrete_system}. Note that an analogous discrete model was introduced in Ref.~\cite{JarzynskiMazonka1999}. In contrast to thorough quantitative analysis of Ref.~\cite{JarzynskiMazonka1999}, here we focus on qualitative discussion of the discrete model with the main aim to understand the basic working principle of the continuous model and in particular appearance of the circulation of the probability current.

%%%%%%%%%%%%%%%%%%%%%%%%%%%%%%%%%%%%%%%%%%%%%%%%%%%%%%%%%%%%%%%%%%%%%%%%%%%%%%%%
%%%%%%%%%%%%%%%%%%%%%%%%%%%%%%%%%%%%%%%%%%%%%%%%%%%%%%%%%%%%%%%%%%%%%%%%%%%%%%%%

In the sketch~\ref{fig:discrete_system} we show one cell of the periodic energy landscape of the discrete ratchet. The red (blue) arrows depict transitions between the discrete states caused by heat exchange with reservoir at the temperature $T_y$ ($T_x$). We assume that the energies of the individual microstates correspond to their vertical position in the sketch, i.e., $\epsilon_1=\epsilon_2=\epsilon_3 < \epsilon_4 < \epsilon_5 < \epsilon_6$. The discrete ratchet thus represents the roughest possible simplification of the complex two-dimensional model discussed in this paper: the transitions between the lower energy levels correspond to the force-free diffusion in the $x$ direction near the channel center ($y = 0$), while the transitions between the upper levels correspond to the diffusion in the $x$ direction at some fixed nonzero $y$, where the particle experiences the asymmetric potential. The sites $1$ and $4$ stand for the $x$ position with the smallest $k(x)$ (widest channel), the sites $3$ and $6$ match the $x$ position with the largest $k(x)$ (narrowest channel).

The main physical assumption imposed on the system is that the transition rates between the individual states fulfill the detailed balance condition. The transition rates in the $x$ direction (blue) satisfy the relation 
\begin{equation}
\frac{r^x_{i\rightarrow j}}{r^x_{j\rightarrow i}} = \exp\!\left(-\frac{ \epsilon_j - \epsilon_i}{T_x}\right).
\label{eq:xDB}
\end{equation}
Similarly we have $r^y_{i\rightarrow j}/r^y_{j\rightarrow i} = \exp[-( \epsilon_j - \epsilon_i)/T_y]$ for the transitions in the $y$ direction (red). These conditions secure that for $T_x = T_y$ the system reaches thermal equilibrium state $\pi_i \propto \exp[- \epsilon_i/T_x]$ with vanishing microscopic probability currents, $\pi_i r_{i\rightarrow j} - \pi_j r_{j\rightarrow i} = 0$.

Let us now assume that the system is initially in thermal equilibrium with $T_x=T_y$ and we slightly increase the temperature $T_x$ (leaving $T_y$ unaltered). Then the detailed balance condition~(\ref{eq:xDB}) implies that the ratio of transition rates for going from lower 
to upper states in the $x$ direction 
to the corresponding rates for going back will be increased as compared to the equilibrium situation. Raising $T_x$ thus leads to positive uphill probability currents in the $x$ direction, the heat flows to the system from the longitudinal bath. In our discrete model, the probability will flow from the state $4$ to the states $5$ and $6$ and from the state $5$ to the state $6$. Meanwhile, the exit rate from state $6$ in the $y$ direction will be the same as in equilibrium and thus the occupation of this state will become larger than in equilibrium once a new stationary occupation of the energy levels consistent with the new reservoir temperatures and non-zero microscopic probability currents will be established. In the 2D ratchet model we observe similar behavior: for $T_x > T_y$ the probability density for position develops global maximum at the $x$ position where the channel is narrowest (see Fig.~\ref{fig:JTxvTy}). The fact that we get microscopic probability currents uphill in the potential landscape together with the continuity of the probability current implies that in our discrete model the probability current will circulate in two circuits: the clockwise circuit $3\rightarrow 2 \rightarrow 1 \rightarrow 4 \rightarrow 5 \rightarrow6 \rightarrow 3$ and the counter-clockwise circuit  $3\rightarrow 1 \rightarrow 4 \rightarrow 6$. Similar circulation of the probability current is found also in the 2D ratchet (see Fig.~\ref{fig:JTxvTy}).

Let us now consider the opposite situation $T_x<T_y$. In analogy with the above reasoning the detailed balance condition (\ref{eq:xDB}) leads to increased downhill rates and decreased uphill transition rates in the $x$ direction with respect to the equilibrium  situation $T_x=T_y$. Probability will thus flow both from left and from right to the state $4$, the heat flows to the system from the transversal bath. Once the system reaches a new steady state the occupation probability of the state $4$ will be larger than the previous equilibrium one. An analogy of this behavior occurs also in the 2D model where the probability density for position develops global maximum at the $x$ position where the channel is widest which can even split in $y$-direction into two global maxima positioned outside the minimum of the potential energy landscape (see Fig.~\ref{fig:JTxmTy}). Again the probability current will circulate in two circuits: the counter-clockwise circuit $3\rightarrow 2 \rightarrow 1 \rightarrow 4 \rightarrow 5 \rightarrow6 \rightarrow 3$ and the clockwise circuit  $3\rightarrow 1 \rightarrow 4 \rightarrow 6$. Also this behavior mimics the circulation of probability currents emerging in the 2D ratchet (see Fig.~\ref{fig:JTxmTy}).

The direction of the global mean probability current can be determined from the following consideration. As we have discussed above, for $T_x>T_y$ the particle
will on average move uphill in the $x$ direction. Both in the discrete and in the two-dimensional model the energy landscape is of the sawtooth type: at some $x$ moving uphill in the $x$ direction corresponds to the current to the right and vice versa for other $x$ positions. However, for the discrete energy landscape of Fig.~\ref{fig:discrete_system} (and also for the potential used in the 2D model), there are less $x$ positions where the probability would flow to the left than to the right (also the probability to move against smaller energy difference is larger) and thus we obtain global mean probability current to the right in accord with Fig.~\ref{fig:TyTx}. Similar reasoning explains why the global probability current in the system is for $T_x<T_y$ directed to the left (see Fig~\ref{fig:TxTy}). 

To close this section let us note that the above reasoning based solely on the detailed balance condition (\ref{eq:xDB}) and general characteristics of the problem (different temperatures in $x$ and $y$ directions and shape of the potential) can be expected to give correct results only in the vicinity of thermal equilibrium, i.e. for small $|T_x-T_y|$. For larger temperature differences the current direction is determined by the detailed form of the transition rates. For example for the exponential rates $r^{x,y}_{i\rightarrow j} = \exp[-(\epsilon_j-\epsilon_i)/(2T_{x,y})]$ the discrete model yields only one current reversal so our close-to-equilibrium analysis gives correct current direction for all temperatures. On the other hand, for the transition rates of the form $r^{x,y}_{i\rightarrow j} = T_{x,y} \exp[-(\epsilon_j-\epsilon_i)/(2T_{x,y})]$ one finds that the mean probability current changes its sign twice and the above reasoning gives the right current direction for small temperature differences only. Finally, for the specific potential (\ref{potentialGEN}), the dynamics of the 2D ratchet is such that the close to equilibrium analysis always gives the correct direction of the current.

%%%%%%%%%%%%%%%%%%%%%%%%%%%%%%%%%%%%%%%%%%%%%%%%%%%%%%%%%%%%%%%%%%%%%%%%%%%%%%%%%%%
%%%%%%%%%%%%%%%%%%%%%%%%%%%%%%%%%%%%%%%%%%%%%%%%%%%%%%%%%%%%%%%%%%%%%%%%%%%%%%%%%%%
\section{Conclusions and outlooks}
\label{sec:conclusions}

The ratchet and pawl mechanism inspired by Feynman's famous thought experiment can be successfully analyzed by a generalization of the Fick-Jacobs theory. The generalization applies far from thermal equilibrium ($T_x\neq T_y$) and captures a fully nonlinear response even for large temperature differences. The mean velocity of rotating wheels and the mean heat current between the reservoirs,  Eqs.~(\ref{eq:V_and_Q_final}), and hence their ratio (the figure of merit, Eq.~(\ref{eq:eta})), are given in terms of scaling functions that depend on the fraction $T_x/T_y$ only. These functions provide a compact description of the ratchet performance in its different working regimes.

The theory predicts a complex behavior of the probability current within the potential unit cell. Further numerical analysis reveals that the ratchet effect is closely related to the circulation of the probability current. The asymmetry of the potential rectifies these vortices and thus the ratchet effect (directed motion) appears. The vortices themselves result from coupling each degree of freedom (wheels and pawl) to heat baths at different temperatures. Their origin is, however, far from being well understood. It is an intriguing open question for a further research to reveal the role of circulating currents in creation of directed transport, their connection with the shape of the potential and with local heat currents. We have also shown that it may be helpful to built an intuition studying discrete systems, which we have used to determine direction of the directed motion on physical grounds.

From a general perspective, the present model and the developed theory offer a rare opportunity to discuss and test laws of irreversible thermodynamics far from thermal equilibrium \cite{Hondou1998, RoeckMaes2007, JarzynskiWojcik2004, Gomez-Marin2006, KomatsuPRL2008, NakagawaKomatsuEPL2006}. Our findings may stimulate further research in this field since the present model serves as a nontrivial example of a strongly nonequilibrium system with known nonlinear response. 
It also could be very interesting to realize the present model experimentally using optical tweezers \cite{MartinezPRE2013, CilibertoPRL2013, Berut2016, Krishnamurthy2016, Martinez2017SoftMat, Martinez2015NatPhys}. The different temperatures $T_x$, $T_y$ can be experimentally realized as described in Refs.~\cite{MartinezPRE2013,Martinez2015NatPhys}. The method described therein can be used to achieve temperature differences up to thousands of Kelvins and the ratchet performance can thus be experimentally investigated effectively in the whole temperature range.

Last but not least, it is worth to apply the presented analytical method to the original Feynman's model with single ratchet wheel and the pawl being pushed against its teeth by a spring \cite{Feynman,Sekimoto1997,Hondou1998,Magnasco1998}. Then, instead of the $y$-symmetric parabolic potential~(\ref{potentialGEN}), one should use an asymmetric potential describing force from the spring and possibly a reflecting boundary condition required when the pawl touches the wheel. This setting is qualitatively similar to the present one, yet different in details (the potential, boundary conditions), which helped us to solve the present model analytically. 
Finally, let us emphasize importance of the spring for the heat transfer between the two reservoirs. To this end, we note that the potential~(\ref{potentialGEN}) should be understood as the simplest model of a ``soft'' repulsion between the pawl and ratchet teeth. It cannot be replaced by a pure elastic hard-wall repulsion without loss of the ratchet effect. For hard-wall repulsion the potential energy $U(x,y)$ is constant everywhere in the channel and there is {\em no heat flow} between the two reservoirs (the expressions for heat flows (\ref{firstlaw}) contain partial derivatives of the potential $U$). Thus for the hard-wall repulsion the two heat reservoirs  decouple and the system cannot work as a ratchet. In all Brownian models of Feynman's original setting \cite{Feynman,Sekimoto1997,Hondou1998,Magnasco1998} there is a potential interaction between the two degrees of freedom, $x$ and $y$, which (possibly in cooperation with the reflecting boundary) allows for a heat transfers between the reservoirs.

%%%%%%%%%%%%%%%%%%%%%%%%%%%%%%%%%%%%%%%%%%%%%%%%%%%%%%%%%%%%%%%%%%%%%%%%%%%%%%%%%%%%%%%%%%%%%%%%%%%%%%%%%%%%%%%%%%%%%
%%%%%%%%%%%%%%%%%%%%%%%%%%%%%%%%%%%%%%%%%%%%%%%%%%%%%%%%%%%%%%%%%%%%%%%%%%%%%%%%%%%%%%%%%%%%%%%%%%%%%%%%%%%%%%%%%%%%%
\ack
Support of the presented research by Czech Science Foundation (project No.\ 17-06716S) is gratefully acknowledged.

%%%%%%%%%%%%%%%%%%%%%%%%%%%%%%%%%%%%%%%%%%%%%%%%%%%%%%%%%%%%%%%%%%%%%%%%%%%%%%%%%%%%%%%%%%%%%%%%%%%%%%%%%%%%%%%%%%%%%
\section*{References}
\bibliographystyle{unsrt}	% (uses file "plain.bst")
\bibliography{myrefs}

\end{document}